\input{aipcheck}

\documentclass[
    ,final            
  ]
  {aipproc}

\layoutstyle{6x9}

\begin{document}

\title
[Magnetic fields in stars]
{Evolution of magnetic fields in stars across the upper main sequence}


\author{S. Hubrig}{
  address={European Southern Observatory, Casilla 19001, Santiago 19, Chile}
}

\author{M. Sch\"oller}{
  address={European Southern Observatory, Casilla 19001, Santiago 19, Chile}
}

\author{P. North}{
  address={Laboratoire d'Astrophysique de l'Ecole Polytechnique F\'ed\'erale
de Lausanne, Observatoire,
CH-1290~Chavannes-des-Bois, Switzerland}
}

\begin{abstract}
 
To properly understand the physics of 
upper main sequence stars it is particularly important to identify the origin 
of their magnetic fields. Recently, we confirmed that magnetic fields appear in 
Ap stars of mass below 3\,$M_\odot$ only if they have 
already completed at least approximately 30\% of their main-sequence 
lifetime \cite{hu00a,hu05a}.
The absence of stars with 
strong magnetic fields close to the ZAMS might be seen as an argument against 
the fossil field theories. Here we present the results of our recent magnetic 
survey with FORS\,1 at the VLT in polarimetric mode
of a sample of A, B and Herbig Ae stars with previously undetected magnetic
fields and briefly discuss their significance for our understanding of the
origin of the magnetic fields in intermediate mass stars.

\end{abstract}

\maketitle


\section{Basic data}

A large variety of physical processes occur in the atmospheres of upper main
sequence stars that have not yet been fully incorporated into stellar models,
or even securely identified and understood. These processes include 
convection,
turbulence, meridional circulation currents, diffusion of trace elements within
the dominant hydrogen plasma, and mass loss through stellar winds.
The chemically peculiar stars (Ap and Bp stars) play a key role in our 
efforts to understand the
relevant physics, since it is in these stars that the effects of the various
processes acting below, in and above the stellar atmospheres are most clearly
visible. Ap and Bp stars 
are main-sequence A and B stars in the spectra of which the lines of some
elements are abnormally strong (e.g., Si, Sr, rare earths) or weak (in 
particular, He). They undergo periodic variations of magnitude (in various
photometric bands) and spectral line equivalent widths and the known periods
of variability range from half a day to several decades. 
Among Ap stars, the magnetic chemically peculiar stars are
particularly important.
For a long time, Ap stars were the only non-degenerate stars besides 
the sun in which direct detections of magnetic fields
had been achieved. Today, they still represent a major fraction of the known
magnetic stars. These stars generally have large-scale organized magnetic 
fields that can be diagnosed through observations of circular polarization in 
spectral lines. The unique large-scale organization of the magnetic fields in 
these stars, which in many cases appears to occur essentially under the form 
of a single large dipole located close to the centre of the star, contrasts 
with the magnetic field of late-type stars, which is most probably subdivided 
in a large number of small dipolar elements scattered across the stellar 
surface.
The fact that magnetic fields of Ap stars are more readily observable than
those of any other type of non-degenerate stars makes them a privileged
laboratory for the study of phenomena related to stellar magnetism.

In spite of the importance of the study of magnetic fields for the proper 
understanding of how the observed fields interact with gravitational settling
and radiative levitation of trace elements, with convection, turbulence,
circulation, and winds or accretion to produce remarkably varied and
inhomogeneous surface chemical abundance patterns, these fields have not been
studied for a representative number of chemically peculiar stars.
To properly understand the physics of Ap stars it is particularly important to 
know the origin of magnetic fields in these stars.
Two main streams of thought have been followed: one
according to which the stars have acquired their field at the time of
their formation or early in their evolution (what is currently observed
is then a {\em fossil\/} field), and the other according to which the field
is generated and maintained by a contemporary {\em dynamo\/} at work
inside the star. 
Whether the A stars become magnetic at a certain evolutionary 
state before reaching the zero-age main sequence (ZAMS) or during the core 
hydrogen burning or at the end
of their main-sequence life  requires systematic studies of established
cluster members, binary systems and of field stars with accurate 
Hipparcos parallaxes. Until now, there is no well-established case of any
known Ap star (either member of a nearby moving cluster or supercluster or 
belonging to a binary system) which is not evolved away from the ZAMS
(e.g., \cite{hu91,hu94,wa96}).

Four years ago we presented the results of a study of the evolutionary
state of magnetic Ap stars using Hipparcos data \cite{hu00a}.
We could show that the distribution of magnetic
Ap stars with accurate Hipparcos parallaxes ($\sigma(\pi)/\pi<0.2$) in the H-R 
diagram differs from that of normal stars in the same
temperature range at a high level of significance. Normal A stars occupy
the whole width of the main sequence, without a gap, whereas magnetic stars are 
concentrated towards the centre of the main-sequence band. In particular, it
was found that magnetic fields appear only in stars that have already
completed at least approximately 30 \% of their main-sequence lifetime.
Knowing the position of the magnetic stars in the H-R diagram, it becomes also 
possible
to probe the evolution of magnetic field strength across the main sequence.
Although our study provided new clues, the results presented in our
work were still inconclusive as to the origin of the magnetic fields of 
Ap stars.
Clearly more magnetic field measurements of Ap stars
for which accurate Hipparcos parallaxes were obtained would be needed to progress. 

Using FORS\,1 at the VLT in service mode during the last two years, new longitudinal field
determinations have been obtained for approximately 50 Ap and Bp stars,
including also normal B stars,
HgMn stars, He weak Si stars, PGa stars and Slowly Pulsating B (SPB) stars.
In addition, our sample included three Herbig Ae stars which sometimes 
in the literature are also called Vega-like stars. As potential progenitors of 
the magnetic Ap stars, Herbig Ae stars provide an excellent
opportunity to study the early evolution of magnetic fields in stars of
identical masses.
A detection of magnetic fields in these stars is especially important in
view of our recent results that Ap magnetic stars of mass
below 3\,$M_\odot$ are significantly evolved
and concentrated towards the centre of the main-sequence band, and
practically no magnetic star of mass below 3\,$M_\odot$ 
can be found close to the zero-age main sequence \cite{hu00a,hu05a}.

FORS\,1 is a multi-mode instrument which is equipped with 
polarization analyzing optics comprising super-achromatic half-wave and 
quarter-wave phase 
retarder plates, and a Wollaston prism with a beam divergence of 22$^{\prime\prime}$ 
in standard resolution mode. For all stars we used the GRISM\,600B in the 
wavelength range 3480--5890\,\AA{} to cover all hydrogen Balmer lines from 
H$_\beta$ to the Balmer jump. The major advantage of using low-resolution 
spectropolarimetry with FORS\,1 is that polarization can be detected in relatively 
fast rotators as we measure the field in the hydrogen Balmer lines.
The determination of the mean longitudinal fields
using FORS\,1 is described in detail in \cite{hu04b}.

\section{Results}

\begin{figure}
  \includegraphics[width=0.65\textwidth]{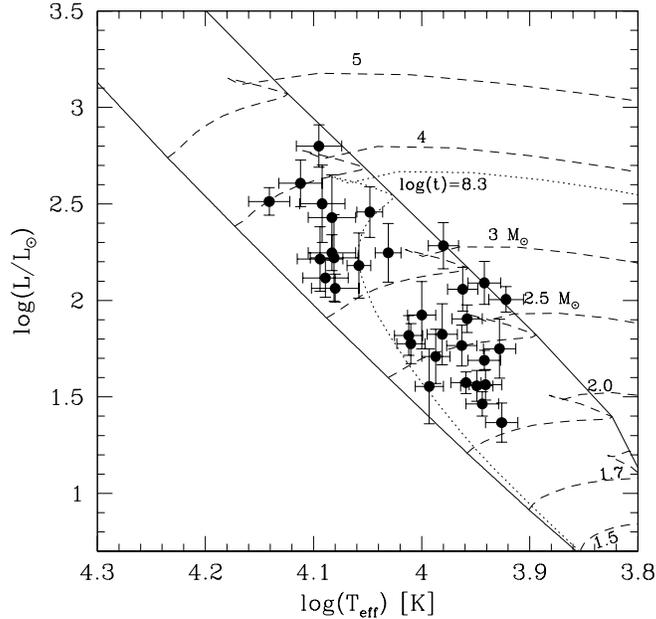}
  \caption{H-R diagram for the sample of magnetic stars with
mean longitudinal fields measured with FORS\,1.}
\end{figure}

The distribution of Ap stars with longitudinal fields measured with
FORS\,1 in the H-R diagram shown in Fig.\,1 confirms our previous finding 
that magnetic stars of mass below 3\,$M_\odot$ are only 
rarely found close to the zero-age main sequence, supporting the view that magnetic
Ap stars are observed only in a restricted range of evolutionary states.
The majority of the rotational periods of the studied stars fall between 2 and 4 
days, 
and there is no indication that the distribution of these stars in the H-R diagram is
different from that of very slowly rotating magnetic stars \cite{hu05a}.

Magnetic stars of higher mass seem to fill the whole width of the 
main-sequence band. 
Our results also show that stronger magnetic fields tend to be found in hotter, 
younger (in terms of the elapsed fraction of main-sequence life) and more massive 
stars.
Hubrig et al.\ \cite{hu00a} have 
already reported about the existence of such a trend in their study of the 
evolutionary state of magnetic Ap stars. 

\begin{figure}
  \includegraphics[width=0.7\textwidth]{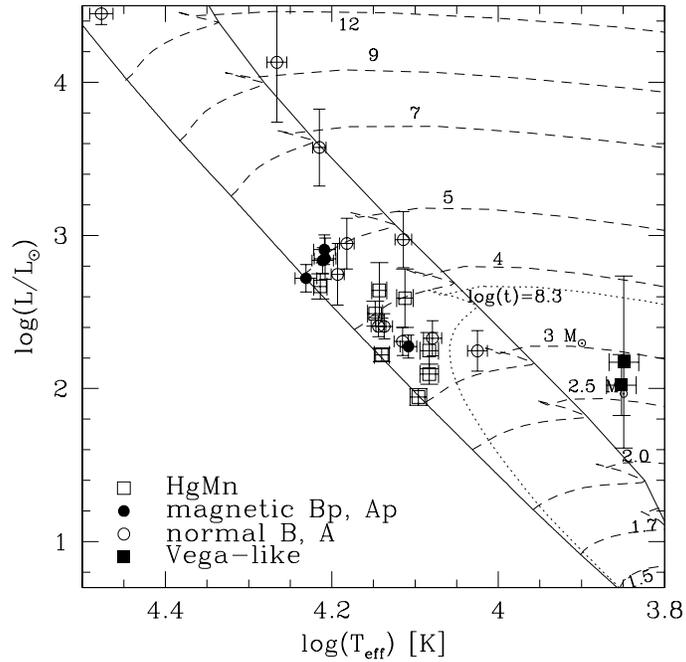}
  \caption{The position of various groups including normal B stars,
HgMn stars, He weak Si stars, SPB stars and two Vega-like stars in the H-R diagram.}
\end{figure}

For the first time a mean longitudinal magnetic field at a level higher
than 3\,$\sigma$ has been detected in one normal B star, two HgMn stars
and four SPB stars \cite{hu05b}.
The position of the studied stars in the H-R diagram is shown in Fig.\,2.
Normal B, HgMn and SPB stars are usually regarded as non-magnetic stars.
The only detection of a magnetic field in an SPB star ($\zeta$~Cas)
has been presented by Neiner et al.\ \cite{Ne03}.
However, the role that magnetic fields
play in the understanding of pulsational properties of SPB stars is still
unclear and further observations are needed to look for possible relations
between magnetic field and pulsation patterns.
It may be an essential clue for the understanding of the origin of the
chemical anomalies of HgMn stars that many stars of this peculiarity
type are very young and are located on the ZAMS or close to it. Previous
searches for magnetic fields in HgMn stars had shown that these stars,
unlike classical Ap stars, do not have large-scale organized fields
detectable through polarimetry. However, we were able to detect longitudinal 
fields of the order of a few hundred Gauss in two HgMn stars.
We also detected a magnetic field at 4.2\,$\sigma$ level in the
normal B-type star HD\,179761.
The intriguing discovery of mean
longitudinal magnetic fields of the order of a few hundred Gauss in a sample
of so-called "non-magnetic" stars rises a fundamental question about the
possible ubiquitous presence of a magnetic field in upper main sequence stars.
The structure of the field in these stars must be, however,
sufficiently tangled so that it does not produce a strong net observable
circular polarization signature.

\begin{figure}
  \includegraphics[width=0.55\textwidth]{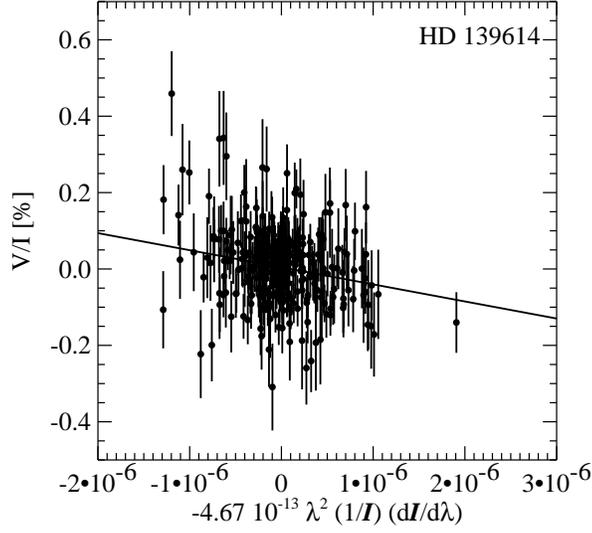}
  \caption{Regression detection of a $-450\pm93$\,G magnetic field in the Vega-like 
star HD\,139614.}
\end{figure}

The observations of the three Herbig Ae stars reveal a definite
longitudinal  magnetic field
in the star HD\,139614 at 4.8\,$\sigma$ level:
$\left<{\cal B}_z\right>$=$-$$450\pm93$\,G \cite{hu04a}.
This is the largest
magnetic field ever diagnosed for a Herbig Ae star. It is diagnosed from the slope 
of a linear regression of $V/I$ versus the quantity
$-g_{\rm eff} \Delta\lambda_z \lambda^2 \frac{1}{I} \frac{{\mathrm d}I}{{\mathrm d
}\lambda} \left<B_z\right> + V_0/I_0$ (Fig\,3).
A hint of a magnetic field is found in the other two stars,
HD\,144432 and HD\,144668, for which
magnetic fields are measured at the $\sim$1.6\,$\sigma$ and $\sim$2.5\,$\sigma$ 
level respectively.
Although magnetic fields are believed to play a crucial role in controlling
accretion onto, and winds from, Herbig Ae stars, contrary to the advance
achieved in magnetic studies of T\,Tauri stars, there is still no 
observational evidence demonstrating the strength, extent, and geometry
of their magnetic fields. To properly assess the role of magnetic 
fields in the star formation process it is 
important to carry out magnetic field studies of a large sample of 
Herbig Ae stars and to try to establish the magnetic field strength and the 
magnetic field geometry by monitoring over several rotation cycles. 

\section{Discussion}

Our recent studies of magnetic fields in stars across the upper main sequence
clearly show that magnetic fields become observable in Ap stars only after
they completed a significant fraction of their life on the main sequence.
By contrast, the magnetic stars of higher mass seem to fill the whole width of the 
main-sequence band. However, we should note that the position of Bp stars in the
H-R diagram is less certain since their effective temperatures 
derived from photometry are not in good agreement with the spectral 
classification \cite{hu00a}. 
Because of the extremely anomalous energy distribution and large variations of 
their spectra, the calibration of the photometric temperature indicators are
frequently questioned. The goal of our future work is to try to resolve these
inconsistencies by detailed spectroscopic studies of these stars.

In Fig.\,1 the isochrone log\,$t$ = 8.3 defines in the H-R diagram the early evolution
envelope of the magnetic Ap star region \cite{hu00a,hu05a}.
These results are in agreement with
the results of the studies of spectroscopic binary systems (e.g., \cite{ca02})
and of the study of Ap stars in the 
cluster NGC~2516 with an age of log\,$t$ = 8.2 $\pm$ 0.1 \cite{ba03}.
Regarding the evolution of magnetic field strength with time, it seems that the 
strongest magnetic fields appear in the middle of the main-sequence band.

The origin of magnetic fields in Ap and Bp stars has been the subject of a long 
debate, which is not closed yet.
The observation that only a rather small portion
of A and B stars are magnetic poses a challenge for both fossil-field
and dynamo theories. The absence of stars with strong magnetic fields 
close to the ZAMS, no observable angular momentum loss before arrival of the stars
on the main sequence \cite{hu00b},
and the absence of strong magnetic fields in Herbig Ae and Be stars
\cite{hu04a} might be seen as an argument 
against the fossil-field theories. 
The competing-dynamo theory proposes that the magnetic field is generated by a 
turbulent dynamo operating in the star's convective core. As long as it was accepted 
that strong magnetic fields are observable at all evolutionary states from the 
ZAMS to the terminal-age main sequence (TAMS), one of the difficulties for the dynamo theory 
was to explain how the field reaches the stellar surface in the rather short time available 
before the arrival of the star on the main sequence. In view of the result 
presented here that magnetic fields become observable only after completion 
of a significant fraction of their main-sequence life it would be interesting to 
reconsider the timescales involved in the possible processes of transport 
of core-generated fields to the surface. 

In summary, although our data provide new clues, the observational results
presented in this work are still inconclusive as to the origin of the
magnetic fields of Ap stars.
The study of the magnetic field geometry in stars of different ages 
and with different rotation rates should provide important additional information 
for testing theoretical predictions. 
 
%
%



\end{document}